\documentclass[a4paper,twoside]{article}
\usepackage{amsmath,amssymb,amsfonts}
\usepackage[
bookmarks=true,
pdfborder={0 0 0},
pdftitle={Chaotic Root-Finding for a Small Class of Polynomials},
pdfauthor={Max Little}
]{hyperref}

\begin{document}

\title{%
\Huge%
\raggedright%
{\bf Chaotic Root-Finding for a Small Class of Polynomials}}
\author{MAX LITTLE$\dagger$ and DANIEL HEESCH$\ddagger$}
\date{}
\maketitle

\noindent
$\dagger${\it Maths Institute, 24-29 St Giles, Oxford University, Oxford OX1 3LB, UK.
Tel: +44 (0) 1865 273525, Fax: +44 (0) 1865 273583, Email: littlem@maths.ox.ac.uk}\\[0.2cm]
$\ddagger${\it Department of Computing, Imperial College London, London SW7 2AZ, UK.
Tel: +44 (0) 20 7594 8298, Fax: +44 (0) 20 7581 8024, Email: dh500@doc.ic.ac.uk}\\

\begin{abstract}
\noindent
In this paper we present a new closed-form solution to a chaotic difference equation,
$y_{n+1} = a_2 y_{n}^2 + a_1 y_{n} + a_0$ with coefficient $a_0 = (a_1 - 4)(a_1 + 2) / (4 a_2)$,
and using this solution, show how corresponding exact roots to a special set of related polynomials
of order $2^p, p \in \mathbb{N}$ with two independent parameters can be generated, for any $p$.
\end{abstract}
{\it AMS Classification Codes:} 39A05, 37E05, 37D45, 34K28\\

\section{{\bf Introduction}}
\parindent 0pt
The difference equation $y_{n+1} = f(y_n), n=0,1,2\dots$ where
\begin{equation}\label{system}
f(y) = a_2 y^2 + a_1 y + \frac{(a_1 - 4)(a_1 + 2)}{4 a_2}
\end{equation}

and $a_1, a_2 \in \mathbb{C},a_2 \not = 0$ has the exact, general solution
\begin{equation}\label{gensol}
y_n(\omega) = \frac{1}{a_2}\left(2 \cos(\omega 2^n) - \frac{a_1}{2}\right) \mbox{.}
\end{equation}

We discovered this solution from the solution presented in \cite{whittaker-91} by
reparametrising with two new variables as, for example, $y_{n} = c \cos(\omega 2^n) + d$.
We then inserted this into the difference equation and equated powers of
$\cos(\omega 2^n)$, obtaining that $c = {2 / a_2}$ and $d = -{{a_1} / {2 a_2}}$.
We were also interested to notice that the fixed-point problem for \eqref{system} is also a
root-finding problem for the polynomial:
\begin{equation}\label{poly}
f^{p}(y) - y = 0
\end{equation}

There is a result sometimes known as {\it Abel's impossibility theorem} which states that there are no
exact expressions for finding the roots of general polynomials of order greater than four in terms of a
finite number of elementary operations. Therefore in general we would be forced to use an iterative root-finding method
for this polynomial (see \cite{householder-70} and \cite{press-92} for example). These methods are typically
difference equations, for example the method of Newton-Raphson iteration for solving the equation
$F(y) = 0$ is, given a close guess $y_0$:
\begin{equation}
\label{newton}
y_{n+1} = y_n - \frac{F(y_n)}{F'(y_n)}
\end{equation}

The main requirement of this method is convergence to some fixed point. However, a problem sometimes
arises in that the system \eqref{newton} may well oscillate, or in some cases behave chaotically. Much progress
has been made by numerical analysts in proving convergence and
inventing better methods that are stable even for bad initial guess values, see for example \cite{isaacson-94}.
Some beautiful methods have been devised that combine numerics with topology. They make it possible to find
approximate solutions of a given system by a continuous deformation of the solutions of a related one that
is exactly solvable, guaranteeing global convergence \cite{allgower-93}.

However, in this case, as in other special cases, we can find the roots of the polynomial \eqref{poly} exactly, without
the need for iteration. The rest of the paper shows how to do this explicitly.

\section{{\bf Periodic Orbits Are Roots of the Polynomial}}
The impossibility theorem is a general one: there are rare and special cases where it does not hold, and this
paper presents another of these special cases, using a result from chaotic dynamics. The difference system
\eqref{system} has two fixed points. It also has a countable infinity
of periodic orbits of all cycle lengths. Rearranging the periodic orbit equation gives equation \eqref{poly}
which is a polynomial equation of order $2^p$, the first two of which are:

\[ a_2 y^2 + (a_1 - 1) y + \frac{(a_1 - 4)(a_1 + 2)}{4 a_2} = 0 \]

and 

\[ {a_2}^3 y^4 + 2 {a_2}^2 a_1 y^3 + a_2 \left(\frac{3}{2} {a_1}^2 - 4 \right) y^2 + \]
\[ \left(\frac{1}{2} {a_1}^3 - 4 a_1 - 1 \right) y + \frac{1}{a_2}\left(\frac{1}{16} {a_1}^4 - {a_1}^2 + \frac{1}{2} a_1 + 2\right) = 0 \mbox{.} \]

Thus, in finding the periodic points of the equation $f^{p}(y) = y$, we also find exact
solutions to the polynomial equations \eqref{poly}. We shall now show the technical details
involved in finding these solutions.

Since the cosine function is 
bounded, so too is the solution \eqref{gensol}. In addition, as $n$ increases, the binary expansion of the expression
$\omega 2^n$ shifts successively leftward, as described in \cite{whittaker-91}, and many other texts on
chaotic dynamical systems (see for example \cite{devaney-89} or \cite{davies-99}).
A key result is that if the binary digit expansion of $\omega / {2 \pi}$ is periodic, then so too is the
behaviour of the solution. Thus, finding the $\omega/2\pi$ that have periodic binary expansions
leads us to the periodic points of $f^{p}(y) = y$, which in turn can be interpreted as the roots of
the $2^p$ order polynomial.

For the purposes of this paper, therefore, only values of
\begin{equation*}
\phi = \omega / {2 \pi}
\end{equation*}
that have periodic binary digit expansions are relevant to us. Since irrational numbers are not
periodic, we must choose $\phi$ rational, i.e. we want $\phi = k / L$
with $k, L \in \mathbb{N}$ such that $\phi$ has a periodic binary digit expansion,
with period $p=\log_2(m)$, where $m$ is the order of the polynomial for which we wish to find the roots.

\section{{\bf Constructing $\phi$}}
Rational fractions with periodic binary digit expansions with periodicity $p$
correspond to the following expressions:
\begin{equation*}
\phi = \sum _{m=1}^{\infty}k (2^p)^{-m} = \frac{k}{2^p - 1} \mbox{.}
\end{equation*}
As mentioned in the previous section, the periodicity of the binary expansion of $\phi$ implies periodicity of the solutions
to \eqref{poly}. To see this, note that:
\begin{eqnarray*}
2^p \phi &=& 2^p \left( \sum _{m=1}^{\infty}k (2^p)^{-m} \right)\\
&=& \sum _{m=1}^{\infty}(2^p) k (2^p)^{-m} = \sum _{m=1}^{\infty}k (2^p)^{(-m+1)}\\
&=& \sum _{m=0}^{\infty}k (2^p)^{-m} = k + \sum _{m=1}^{\infty} k (2^p)^{-m}\\
&\equiv& \sum _{m=1}^{\infty}k (2^p)^{-m} \equiv \phi (\mbox{mod}\; 1)
\end{eqnarray*}

Therefore, $2^p \phi \equiv \phi (\mbox{mod} \; 1)$, which implies, since $\omega = 2\pi \phi$,
that $2^p \omega \equiv \omega (\mbox{mod}\;\; 2\pi)$, and we reach the conclusion that
$\cos \left(2^p \omega \right) = \cos(\omega)$, as required.

There is, however, a minor complication to this scheme due to the symmetry of the cosine function
about $\pi$ in the general solution \eqref{gensol}. This symmetry
implies that values of $\omega$ that lie equidistant from $\pi$ produce the same
solution. Then
\begin{equation*}
\left|\omega - \pi \right| = \left|\phi 2\pi - \pi \right| =
\pi \left|\frac{2k}{2^p - 1} - 1 \right| =
\frac{\pi}{2^p - 1}\left|(2k + 1) - 2^p \right|
\end{equation*}

must be unique for all choices of $k$. However,

\begin{equation}\label{symmetric}
|(2n + 1) - 2^p| = |2^p - (2n + 1)|\mbox{,}
\end{equation}
so that $k = m$ and $k = 2^p - (m + 1)$ for  $m = 0,1 \dots 2^p - 1$ lead to symmetrically
identical solutions to the polynomial. Therefore, only $k = 0,1 \dots 2^{p-1} - 1$
give the unique required solutions.\\
\\%
The consequence of this is that in order to find all solutions of the $2^p$ order polynomial,
we must seek solutions to the polynomial of order $2^{2p}$ instead, since all periodic points
of $f^{p}(y) = y$ also satisfy $f^{2p}(y) = y$. We therefore set $L = 2^{2p} - 1$. However,
the converse is not true: not all the periodic points of $f^{2p}(y) = y$ are periodic points
of $f^{p}(y) = y$. Therefore, in enumerating the solutions of the polynomial, $\phi$
must satisfy two criteria:

(i) As mentioned above, the symmetry of the $\cos$ function implies that $\phi$ must
be less than $\frac{1}{2}$, which in turn implies that $k < 2^{2p - 1}$\\
(ii) $\phi$ must either have a periodic binary digit expansion with period $p$, or period
$2p$ but have $\omega$ symmetric about $\pi$ under the iterative shift of $p$ digits,
i.e. $\phi$ and $\phi 2^p$ lie equidistant from $\pi$.

Therefore, to construct and enumerate all $\phi$ with binary digit expansions of
period $p$ when embedded in a sequence of $2p$, we set:
\begin{equation*}
k = m + m 2^p = m (2^p + 1)
\end{equation*}
for $m = 0,1 \dots 2^{p-1} - 1$. We thus enumerate half of the required solutions. Secondly,
from \eqref{symmetric}, to find all the symmetric $\phi$ with binary digit expansions of
period $2p$, we then choose:
\begin{equation*}
k = m 2^p - m = m (2^p - 1)
\end{equation*}
for $m = 1, 2 \dots 2^{p-1}$, and we have thereby enumerated the remaining solutions.

\section{{\bf A High-Order Example}}
Here we demonstrate an application of the method to the order eight polynomial $f^3(y) - y = 0$.
The following expressions for the coefficients in ascending order of $y$ are:
\begin{eqnarray*}
y^0 &:& \frac{1}{a_2}\left(2 - \frac{1}{2} a_1 - 4 a_1^2 + \frac{5}{4} a_1^4
- \frac{1}{8} a_1^6 + \frac{1}{256} a_1^8 \right)\\
y^1 &:& -1 - 16 a_1 + 10 a_1^3 - \frac{3}{2} a_1^5 + \frac{1}{16} a_1^7\\
y^2 &:& a_2 \left(-16 + 30 a_1^2 - \frac{15}{2} a_1^4 + \frac{7}{16} a_1^6 \right)\\
y^3 &:& a_2^2 \left( 40 a_1 - 20 a_1^3 + \frac{7}{4} a_1^5 \right)\\
y^4 &:& a_2^3 \left( 20 - 30 a_1^2 + \frac{35}{8} a_1^4 \right)\\
y^5 &:& a_2^4 \left( -24 a_1 + 7 a_1^3 \right)\\
y^6 &:& a_2^5 \left( -8 + 7 a_1^2 \right)\\
y^7 &:& 4a_1 a_2^6\\
y^8 &:& a_2^7
\end{eqnarray*}
We use the LaGuerre root-finding method \cite{press-92}
with $a_2 = -1 + i, \ a_1 = 2 - i$, and we find that numerical solutions to the
polynomial accurate to four decimal places are:
\begin{eqnarray*}
y & = & \{-0.25 - 0.75i, -0.016 - 0.516i, 0.1265 - 0.3735i, 0.5764 + 0.0764i, \\
& & 0.97252 + 0.4725i, 1.25 + 0.75i, 1.651 + 1.151i, 1.6897 + 1.1897i \}
\end{eqnarray*}
For this polynomial, $p = 3$ and so $L = 2^{2p} - 1 = 63$. Next, we enumerate the solutions, firstly for
$m = 0,1 \dots 2^{p-1} - 1 = 3$. Therefore,
\begin{eqnarray*}
k & = & m (2^p + 1) = 0, 9, 18, 27\\
\phi_m & = & \left\{ \frac{0}{63}, \frac{9}{63}, \frac{18}{63}, \frac{27}{63} \right\}
= \left\{ 0, \frac{1}{7}, \frac{2}{7}, \frac{3}{7} \right\} \\
\mbox{or in binary notation}\\
\phi_m & = & \left\{ 0.{\overline{000000}}, 0.{\overline{001001}}, 0.\overline{010010}, 0.\overline{011011} \right\} \\
\end{eqnarray*}
where for example $0.\overline{010010}$ indicates the infinite binary digit repetition of the sequence
010010.

From these values of $\phi_m$ we then calculate the first half set of solutions, here given to eight
decimal places:
\begin{eqnarray*}
y_m & = & \frac{1}{a_2}\left(2 \cos(\phi_m 2\pi ) - \frac{a_1}{2}\right)\\
& = & \{-0.25 - 0.75i, 0.1265102 - 0.3734898i, 0.97252093 + 0.47252093i,\\
& & 1.65096887 + 1.15096887i \}
\end{eqnarray*}
Secondly, we enumerate the symmetric set for $m = 1, 2 \dots 2^{p-1}$:
\begin{eqnarray*}
k & = & m (2^p - 1) = \{ 7, 14, 21, 28 \}\\
\phi_m & = & \left\{ \frac{7}{63}, \frac{14}{63}, \frac{21}{63}, \frac{28}{63} \right\}
= \left\{ \frac{1}{9}, \frac{2}{9}, \frac{1}{3}, \frac{4}{9} \right\} \\
& = & \left\{ 0.{\overline{000111}}, 0.{\overline{001110}}, 0.\overline{010101}, 0.\overline{011100} \right\}\end{eqnarray*}
from which we can calculate the second set of solutions, again to eight decimal places:
\begin{eqnarray*}
y_m & = & \frac{1}{a_2}\left(2 \cos(\phi_m 2\pi ) - \frac{a_1}{2}\right)\\
& = & \{ -0.01604444 - 0.51604444i, 0.57635182 + 0.07635182i, 1.25 + 0.75i,\\
& & 1.68969262 + 1.18969262i \}
\end{eqnarray*}

\section{{\bf Conclusions}}
By finding a general solution to a discrete, chaotic system and interpreting 
a problem of finding roots of a high-order polynomial as the fixed point equation
for that chaotic system, we have shown how to obtain exact roots to the polynomial.
We then compared this with the results of a numerical root-finding method. Of course
we have not presented a {\it general} root-finding method, but we find it intriguing
to notice that iterative root-finding methods are often difference equations in their
own right, for which general solutions of particular cases may well be known.

\bibliographystyle{unsrt}
\addcontentsline{toc}{section}{\protect\numberline{}{References}}
\bibliography{chaos}

\end{document}